\documentclass[11pt]{article}

\usepackage[final]{acl}

\usepackage{times}
\usepackage{latexsym}

\usepackage[T1]{fontenc}

\usepackage[utf8]{inputenc}

\usepackage{microtype}

\usepackage{inconsolata}

\usepackage{graphicx}

\usepackage{fontawesome5}
\usepackage{hyperref}
\usepackage{url}
\usepackage[utf8]{inputenc} 
\usepackage[T1]{fontenc}    
\usepackage{hyperref}       
\usepackage{url}            
\usepackage{booktabs}       
\usepackage{amsfonts}       
\usepackage{nicefrac}       
\usepackage{microtype}      
\usepackage{xcolor}         
\usepackage{amsmath}
\usepackage{amssymb}

\usepackage{amsfonts}
\usepackage{mathtools}
\usepackage{amsthm}
\usepackage{multirow}
\usepackage{booktabs}
\usepackage{amsfonts,pifont,tabularx}
\usepackage{algorithm}
\usepackage{algpseudocode}
\usepackage{multirow}
\usepackage{tabu}
\usepackage{graphicx}
\usepackage{tabularx}
\usepackage{textcomp}
\usepackage{xcolor}
\usepackage{url}
\usepackage{mdframed}
\usepackage{multirow, makecell}
\PassOptionsToPackage{table,xcdraw,dvipsnames}{xcolor}
\usepackage{colortbl}
\usepackage{soul}
\usepackage{bm}
\usepackage{latexsym}
\usepackage{mathtools}
\usepackage{enumitem}
\usepackage{diagbox}
\usepackage{wrapfig}
\usepackage{tikz}
\usepackage{pgfplots}
\usepgfplotslibrary{groupplots}
\pgfplotsset{compat=newest}
\usepackage{array}
\newcolumntype{?}{!{\vrule width 1pt}}
\usepackage{caption} 
\usepackage{wrapfig}

\usepackage{soul}
\usepackage[normalem]{ulem}
\usepackage{xspace}
\usepackage{lipsum}
\usepackage{caption} 
\usepackage{subcaption} 
\usepackage[most]{tcolorbox}
\usepackage{pifont}

\title{MemoPhishAgent: Memory-Augmented Multi-Modal LLM Agent for Phishing URL Detection}

\author{
Xuan Chen\thanks{This work was done during an internship at Amazon.} \\
Purdue University \\
\texttt{chen4124@purdue.edu} \\
\And
Hao Liu \\
Amazon \\
\texttt{liuahl@amazon.com} \\
\And
Tao Yuan \\
Amazon \\
\texttt{yuanty@amazon.com} \\
\AND
Mehran Kafai \\
Amazon \\
\texttt{mkafai@amazon.com} \\
\And
Piotr Habas \\
Amazon \\
\texttt{phabas@amazon.com} \\
\And
Xiangyu Zhang \\
Purdue University \\
\texttt{xyzhang@purdue.edu} \\
}

\newcommand{\sys}{{\sc MPA}\xspace}

\begin{document}
\maketitle

\begin{abstract}
Traditional phishing website detection relies on static heuristics or reference lists, which lag behind rapidly evolving attacks. 
While recent systems incorporate large language models (LLMs), they are still prompt-based, deterministic pipelines that underutilize reasoning capability.
We present MemoPhishAgent (\sys), a memory-augmented multi-modal LLM agent that dynamically orchestrates phishing-specific tools and leverages episodic memories of past reasoning trajectories to guide decisions on recurring and novel threats.
On two public datasets, \sys outperforms three state-of-the-art (SOTA) baselines, improving recall by 13.6\%.
To better reflect realistic, user-facing phishing detection performance, we further evaluate \sys on a benchmark of real-world suspicious URLs actively crawled from five social media platforms, where it improves recall by 20\%.
Detailed analysis shows episodic memory contributes up to 27\% recall gain without introducing additional computational overhead.
The ablation study confirms the necessity of the agent-based approach compared to prompt-based baselines and validates the effectiveness of our tool design.
Finally, \sys is deployed in production, processing $\sim 60\text{K}$ targeted high-risk URLs weekly, and achieving 91.44\% recall, providing proactive protection for millions of customers.
Together, our results show that combining multi-modal reasoning with episodic memory yields robust phishing detection in realistic user-exposure settings. 
Our implementation is available at \url{https://github.com/XuanChen-xc/MemoPhishAgent.git}.

\end{abstract}

\section{Introduction}
\label{sec:intro}

\begin{figure}[t]
    \centering
     \includegraphics[width=0.48\textwidth]{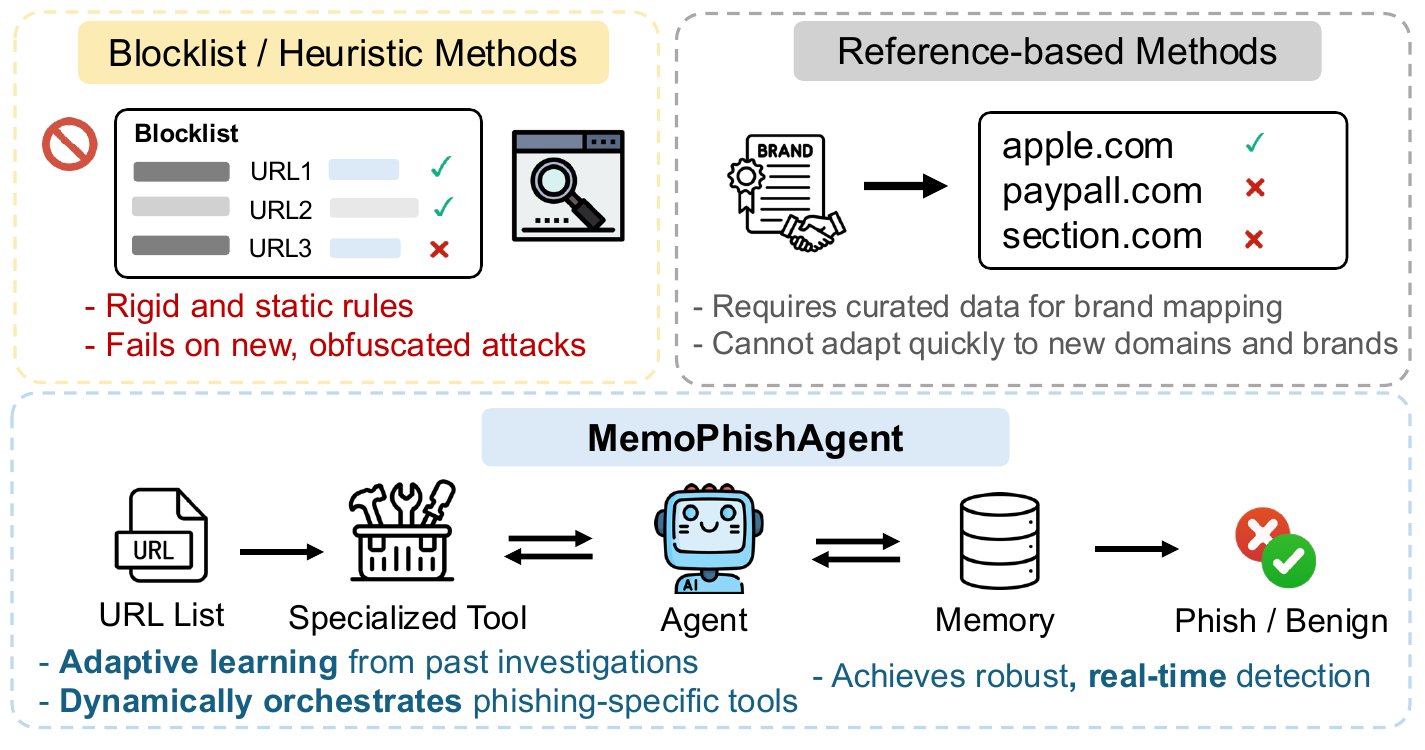}
    \caption{\small Comparison between the previous heuristic and reference-based detection methods and our agent pipeline.}
    \label{fig:comparison}
    \vspace{-3mm}
\end{figure}
Phishing remains a pervasive and costly threat, with attackers continually evolving to evade detection~\citep{insidephish, PhishEye, catchphish, wang2025can}. 
As shown in Fig.~\ref{fig:comparison}, traditional defenses based on static blocklists and handcrafted heuristics can be effective for known attacks, but their coverage quickly degrades when attackers register new domains, obfuscate content, or change infrastructure~\citep{phishtank, Garera2007, Sheng2010, afroz2011phishzoo, zhang2007cantina, Xiang2011}.
Reference-based detectors improve robustness by validating URLs against curated brand-domain mappings, reducing look-alike and typosquatting risks~\citep{li2025phishintel, li2024KnowPhish, Liu2023, liu2022inferring, lin21Phishpedia}.
However, as summarized in Fig.~\ref{fig:comparison}, these mappings are expensive to maintain and can lag behind emerging brands, new subdomains, and fast-moving campaigns, becoming a bottleneck in practice~\cite{kulkarni2024phishing, ji2025evaluating}.

Learning-based approaches further expand coverage by training classifiers on multi-modal signals from URLs, HTML content, and screenshots~\citep{maneriker2021urltran, karim2023phishing, iftikhar2024upadm, li2024state}. 
By feeding these features into classifiers, e.g., gradient-boosted trees or transformers, these methods achieve higher detection coverage and adaptability than rule-based systems.
Nonetheless, they often require extensive feature engineering and retraining to accommodate new phishing tactics, which can be resource-intensive.
To improve reliability, emerging work orchestrates LLMs as agents that coordinate tools via multi-step reasoning~\citep{wang2024automated, li2025phishdebate, nakano2025scamferret} or directly serve as an internal knowledge base~\citep{liu2024less, shen2023hugginggpt}.
Despite better flexibility, they rigidly rely on general-purpose tools, which are not specifically designed for phishing detection.
~\citet{cao2025phishagent} propose a multi-modal LLM-based solution for phishing URL detection. 
They use LLM to interpret tool outputs, which improves reliability, but still restricts adaptive evidence acquisition.
The reasoning is performed under partial evidence, where the most informative next step depends on what has already been observed.
Additionally, memory-less designs fail to leverage historical investigations, limiting both accuracy and efficiency in recurring patterns.

To address these challenges, we propose and deploy MemoPhishAgent, the first multi-modal LLM agent specialized for phishing URL detection that dynamically selects and sequences phishing-specific tools based on the current evidence state, rather than following a pre-defined workflow. 
Inspired by how human experts investigate suspicious sites, \sys is equipped with five specialized, multi-modal tools that collect complementary evidence. 
To improve detection efficiency, we further introduce a novel memory system to store, retrieve, and manage historical reasoning trajectories, and learn from all its past interactions, refining its tool-calling strategies over time. Our contributions can be summarized as follows:

\ding{182} We introduce MemoPhishAgent, \textbf{the first agent-based framework specifically designed for real-world phishing URL detection.}
Unlike prior approaches that struggle under non-stationary threats and real-world constraints, we design multi-modal tools and enable the agent to choose them based on the current evidence rather than following a fixed workflow, improving both detection accuracy and efficiency.

\ding{183} We propose \textbf{a novel episodic memory that stores prior reasoning trajectories and outcomes, and retrieves them as targeted exemplars and heuristics.} It turns past investigations into reusable experience, enabling fast decisions on recurring patterns and better targeted exemplars for hard cases.

\ding{184} We show that \textbf{\sys outperforms SOTA baselines on both static benchmarks and a real-world dataset} crawled from social media platforms. We further demonstrate effectiveness through \textbf{production deployment that protects millions of users.}
Comprehensive ablation studies validate the necessity of our memory system and specialized tools.
\section{Related Work}
\label{sec:rw}
\textbf{Classical and reference-based methods.}
Early phishing URL detectors relied on handcrafted blocklists and heuristics to identify malicious sites~\citep{Garera2007, Sheng2010, zhang2007cantina, Xiang2011, afroz2011phishzoo}.
By matching URLs against known phishing lists~\citep{khonji2013phishing, openphish}, or applying rules based on URL token patterns, and HTML structures, these systems achieve strong precision for known brands. 
However, their dependence on pre-defined rules and lists limited their ability to keep up with evolving phishing tactics.
To address these limitations, reference-based detectors introduced brand-domain mapping~\citep{li2025phishintel, li2024KnowPhish, Liu2023, liu2022inferring, lin21Phishpedia}.
These methods first built a reference list that specifies the brand and the authentic domains.
Given a URL, they extracted the brand information and verified whether the domain aligns with the known set.
If a mismatch occurs, the URL is flagged as phishing.
While this approach added a semantic layer beyond simple pattern matching, it still hinged on an up-to-date brand list, which fails to generalize to new brand subdomains and thus undermines detection coverage.

\noindent \textbf{ML-based methods.}
ML-based approaches \citep{maneriker2021urltran, karim2023phishing, iftikhar2024upadm} detected phishing websites by extracting and analyzing specific features instead of relying on static lists.
The features were derived from the domain pattern, HTML file, and embeddings of the screenshot~\citep{li2024state}.
These multi-modal feature sets were fed into ML or transformer architectures~\citep{maneriker2021urltran} to improve the coverage of various phishing schemes.
However, these methods depend on costly, expert-driven feature engineering, still struggle to adapt quickly to evolving and novel phishing patterns, as we need to retrain our model based on new features.

\noindent \textbf{LLM and agent-based methods.}
These approaches prompt LLMs to inspect HTML, screenshots, or crawled text, leveraging their multimodal understanding and internal knowledge for phishing detection~\citep{liu2024less, koide2023detecting, lee2024multimodal}.
However, standalone LLMs are often deployed in single-pass, prompt-only settings for stability and cost control, which limits evidence gathering and amplifies false positives on edge cases~\citep{ wang2024automated, li2025phishdebate, nakano2025scamferret}. 
To improve reliability, recent work introduces agent-based pipelines that orchestrate tools and multi-step reasoning~\citep{yao2023react, shinn2023reflexion, cao2025phishagent}, enabling richer evidence collection and handling more complex URLs.
Beyond building agent pipelines, recent work also systematically test tool-using LLM agents for intent preservation and safety coverage~\citep{feng2025intentest,chen2026safeaudit}.
Those pipelines are still embedded into deterministic evidence acquisition policies, a tradeoff that improves predictability but makes the overall pipeline less responsive to novel cases.
Moreover, the tools are general-purpose rather than phishing-tailored, leaving substantial room to refine both the evidence acquisition strategy and the tools for real-world detection.

\section{Methodology}
\label{sec:methodology}

\subsection{Problem Setting \& Threat Model}

\textbf{Threat model.} 
Following prior studies~\citep{liu2024less, dynaphish, karim2023phishing, li2025phishintel}, we consider an attacker whose goal is to harvest credentials or personal data, or to induce harmful user actions, by luring victims to attacker-controlled web resources.
The attacker may mimic legitimate brands or use generic lures (e.g., ``free gift cards'') without explicit impersonation to attract clicks and form submissions. 
Attackers control client-side content, including text, images, hidden inputs~\citep{ahammad2022phishing, sanchez2022phishing}, and may use obfuscation, URL shorteners, and multi-hop redirects to evade detectors.
We assume an adaptive adversary who understands common detection strategies and can tune pages to bypass heuristic checks, but cannot tamper with our deployed detection pipeline or modify the agent’s memory once deployed. 

\noindent \textbf{Our scope.}
Our work targets online phishing detection in realistic, user-exposure settings. The detector operates on suspicious URLs surfaced from user-facing channels, e.g., social platforms, messaging, or other online sources, and analyzes live, in-the-wild pages that users could actually visit.
These pages often remain interactive, and may change over time, making the evidence state inherently non-stationary~\citep{li2024KnowPhish, CTINexus}.
This differs from offline pipelines that classify pre-collected artifacts or fixed evidence~\cite{cao2025phishagent}, and motivates our design for adaptive tool use and memory-guided investigation under evolving web content. In practice, candidate URLs can be verified downstream, e.g., via third-party services, so our deployment objective prioritizes high recall to minimize missed phishing URLs, while tolerating a manageable number of false positives.

\begin{figure*}[t]
    \centering
    \includegraphics[width=0.97\textwidth]{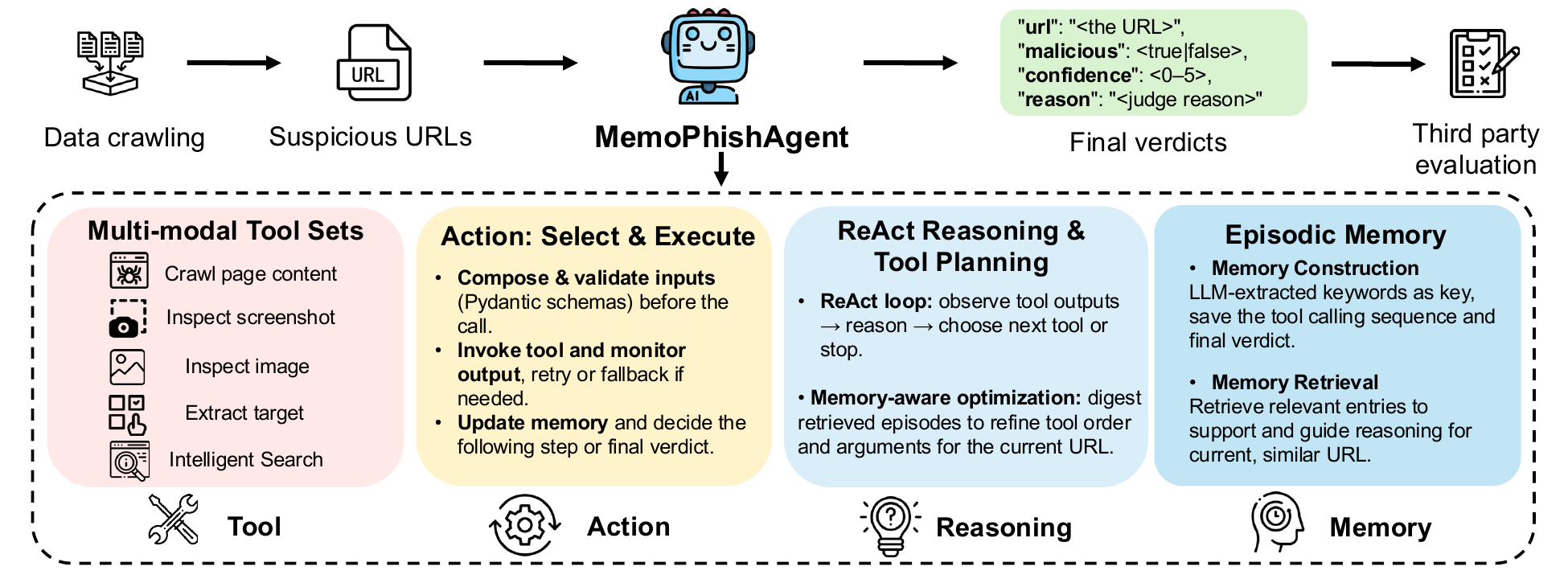}
    \caption{{\small Overview of \sys. It includes optional data crawling to collect suspicious URLs, a toolkit of page crawling, screenshot inspection, target extraction, and intelligent search engine modules, dynamic action selection and execution within a ReAct reasoning loop, and an episodic memory module for planning and retrieval.}}
    \label{fig:overview}
\end{figure*}

\noindent \textbf{Overview.}
Fig.~\ref{fig:overview} presents the end-to-end architecture of \sys.
Starting from a list of suspicious URLs,
each sample is passed to the agent, which dynamically orchestrates five specialized tools to gather evidence, perform multi-step reasoning, and arrive at a ``malicious'' or ``benign'' verdict, with a one-sentence summary as the reason.  
URLs deemed malicious are then submitted to third-party evaluators for ground-truth verification.



In the following sections, we provide motivations and designs for our tools, episodic memory structures, and the memory-aware optimization of tool-calling strategies.

\subsection{Tool Design}
\label{subsec:tools}

We design five specialized tools around three key aspects, enabling complementary multi-modal evidence for robust phishing detection:

\noindent \textbf{Multi-modal evidence (text + vision).}
  Phishing cues appear across HTML semantics and visual appearance, so we provide tools for both:
  \begin{itemize}[leftmargin=*, topsep=1pt, itemsep=1pt, parsep=0pt]
    \item \textbf{Crawl Content:} fetches the page and converts raw, noisy HTML into an LLM-friendly markdown representation.
    This reduces token waste from HTML tags and helps the agent focus on semantic signals.
    \S\ref{appx:robust_crawl} shows that our agent is robust to the choice of crawled content representation.
    \item \textbf{Check Screenshot:} summarizes a full-page screenshot and prompts the LLM to assess whether the page looks suspicious.
    \item \textbf{Check Image:} enables fine-grained visual inspection by isolating salient graphical elements
    for targeted analysis when the full-page view is inconclusive.
  \end{itemize}

\noindent  \textbf{External knowledge for non-stationary threats.}
  As phishing tactics and benign domains evolve quickly, internal LLM knowledge can be stale:
  \begin{itemize}[leftmargin=*, topsep=1pt, itemsep=1pt, parsep=0pt]
    \item \textbf{Intelligent Search:} allows the agent to form evidence-driven search queries (e.g., ``is \textit{epik.com} legitimate?'') and incorporate retrieved results as additional context,
    improving judgments on unfamiliar and emerging patterns.
  \end{itemize}

\noindent  \textbf{Nested attack surfaces.}
  Phishing sites hide the malicious step behind redirects or child pages:
  \begin{itemize}[leftmargin=*, topsep=1pt, itemsep=1pt, parsep=0pt]
    \item \textbf{Extract Targets:} identifies suspicious embedded links or redirect targets and surfaces child URLs for deeper investigation,
    which can then be examined using the other tools.
  \end{itemize}

Together, these tools let the agent adapt evidence collection to the current uncertainty state. 
Implementation details of each tool are included in \S\ref{appx:tech_details}.

\subsection{Episodic Memory System}
\label{subsec:memory}

Equipped with five tools, \sys iteratively selects tools and reasons over evidence under the ReAct~\citep{yao2023react} framework until producing a final verdict. 
Each URL yields a tool-calling trajectory and decision, which we treat as reusable supervision: when similar URLs recur, the agent can retrieve prior investigations to accelerate and stabilize its reasoning. 
We therefore introduce an \emph{episodic memory} module, inspired by retrieval-augmented LLMs and memory systems~\citep{lewis2020retrieval, guu2020realm, graves2016hybrid, zhou2024assistrag, xu2025mem}. 
The memory serves two roles: \textbf{fast recall}, by surfacing similar past trajectories to reuse effective evidence-collection paths, and \textbf{robust consensus}, by aggregating judgments over multiple similar episodes to reduce hallucinations and single-sample bias.

\subsubsection{Memory Construction}
\label{subsubsec:memory_construction}

First, we need an index to judge whether the URLs are similar.
Directly using the raw text or the embeddings of the URL content is inefficient, as raw text often contains lots of redundant information, which distracts the agent when determining similarity. 
As such, we prompt the LLM to distill compact keywords $\{\mathbf{k}_i\}$, capturing core semantics (e.g., ``apple~login'', ``invoice~pdf'', ``wallet~connect''), given the page text and screenshot. 
We embed the resulting keyword set $\mathbf{k}$ with a pre-trained sentence encoder and store it in a vector index along with its \emph{value}: the full tool-calling sequence $\langle t_1,\dots,t_m\rangle$ and the final verdict $\hat{y}\in\{\text{malicious},\text{benign}\}$.  
This design keeps storage lightweight while preserving the entire reasoning trace for later reuse.

\subsubsection{Memory Retrieval and Management}
\label{subsubsec:memory_retrieval}

\textbf{Similarity retrieval.}  
Given a new URL, we extract its keyword set $\mathbf{k}^{\mathrm{new}}$ and retrieve the top-$k$ nearest neighbors $\{\mathbf{k}^{(1)},\dots,\mathbf{k}^{(k)}\}$ by cosine similarity; neighbors above a threshold $\tau$ form the matched set of size $k'$.
When memory is sparse (early in deployment), $k'$ may be zero; as memory grows, retrieval increasingly returns relevant prior episodes.


\noindent \textbf{Memory-aware judgment strategy.}
While retrieved memories provide valuable guidance, they should not dominate the agent's reasoning.
Instead, we aim for them to act as a context to improve detection efficiency by reusing prior interactions. 
We employ a three-tier policy that balances speed and reliability for the retrieved memory:

\begin{itemize}[leftmargin=*, nosep]
  \item \textbf{No match ($k'=0$).} If no neighbor exceeds $\tau$, the agent runs the default \textbf{\emph{full ReAct}} loop, generating a complete reasoning chain from scratch.
  \item \textbf{Partial match ($0<k'<k$).} When a subset of neighbors exceeds $\tau$, retrieved episodes are provided as \textbf{\emph{in-context exemplars}}; the agent reuses historical traces to invoke only the tools needed for additional evidence.
  \item \textbf{Full match ($k'\ge k$).} The agent returns a \textbf{\emph{majority vote}} over stored verdicts of the retrieved episodes.
\end{itemize}

This hierarchical scheme yields fast responses for recurring phishing patterns. It also guides reasoning for partially novel cases and full analysis for unseen attacks.  
Ablation results in Section~\ref{subsec:ablation} validate the contribution of each memory branch. 
In \S\ref{appx:memory_pruning}, we evaluate different memory pruning and forgetting policies, and analyze robustness under imperfect stored episodes. We further break down the performance characteristics of each branch in \S\ref{appx:memory_submodule}.

\section{Experiment}
\label{sec:exp}

\textbf{Datasets.}  We evaluate \sys on two static datasets.
\textbf{TR-OP}~\citep{li2024KnowPhish} is a manually labeled dataset containing 5,000 phishing and 5,000 benign URLs.
Phishing URLs are obtained from OpenPhish~\citep{openphish} and validated between July and December 2023, while benign URLs are randomly drawn from the top 50,000 Tranco domains \citep{pochat2018tranco}.
\textbf{DynaPD}~\citep{Liu2023, liu2024less} is a website-level benchmark containing 6,075 phishing sites, each manually crafted to mimic real-world attacks.
Their 6,075 benign websites are crawled from the Alexa~\citep{alexaweb} top 5,000 to 15,000 websites.


\noindent \textbf{Baselines.} 
We select four representative baselines: PhishLLM~\citep{liu2024less}, PhishIntention~\citep{liu2022inferring}, MLLM~\citep{lee2024multimodal}, and URLTran~\citep{maneriker2021urltran}. 
PhishLLM and PhishIntention improve reference-based methods by using an LLM to extract brand mentions, identify credential-taking intention, and verify brand-domain consistency. 
MLLM first employs a vision-enabled LLM that captures brand signals from screenshots, and another LLM integrates these visual cues with URL characteristics to make a final judgment. 
URLTran is a transformer-based URL-only discriminative classifier fine-tuned on raw URL strings, serving as a strong URL-pattern baseline. 

\noindent \textbf{Setup.}
For implementation details, both the Check Screenshot and Check Image tools use Claude-3-Sonnet for multimodal webpage analysis, while the Intelligent Search tool is backed by Google Search via SerpAPI.
To analyze redirects and embedded links, the Extract Targets tool first enumerates all original links, redirect targets, and embedded links from the current page into a flat list of child URLs.
Each child URL is then examined independently by the same tool suite without recursively re-invoking Extract Targets, so the process is acyclic and terminates deterministically.
The final propagation rule is one-directional: if any redirect target or embedded link is judged malicious, the original URL is also labeled malicious.

\subsection{Results \& Analysis}

\textbf{Effectiveness \& Efficiency.}
In Tab.~\ref{tab:effective}, \sys consistently achieves the best recall and $F_{1}$ score across both datasets compared to baselines.
On TR-OP, it boosts recall $\sim$13\% while also improving ACC and $F_{1}$, indicating that the gain is not merely from over-flagging but from stronger evidence aggregation. 
On DynaPD, \sys raises recall to 93.6\% and achieves the top $F_{1}$, outperforming prompt-only and fixed-pipeline baselines that tend to miss positives. 
We prioritize recall because, in deployment, candidate URLs are verified by third-party services rather than ground-truth labels, which are also not available. Missing true phishing URLs is costlier than reviewing additional candidates. 
Verification also incurs cost, so we aim for an operating point that achieves high recall while keeping the number of reviewed URLs manageable.
Overall, Tab.\ref{tab:effective} indicates that \sys reliably detects phishing websites across diverse benchmark settings.
In Tab.~\ref{tab:effective_full}, on both of TR-OP and DynaPD, URLTran achieves the highest ACC and $F_{1}$, indicating that this conventional benchmark can be handled well by URL-pattern modeling alone.
The recall of URLTran drops to 0.8039 on the real-world dataset SocPhish, which uses URL shorteners and platform-hosted paths (e.g., ``sites.google.com'') that are lexically indistinguishable from benign traffic, while \sys still attains the best recall. 

In Tab.~\ref{tab:latency}, \sys is the fastest method overall, requiring 4.46s per URL.
Because it is impractical to deploy all baselines in our production environment, we run all efficiency experiments locally on a single RTX A6000 GPU and measure end-to-end processing time per URL.
For \sys and MLLM, latency is measured end-to-end including external LLM API time and does not rely on local GPU inference.
For URLTran, which includes a separate training stage, we report the combined training and inference time divided by the total number of URLs; its inference-only latency is 1.07s/URL.
These numbers are also affected by engineering choices such as current LLM API latency and the specific web crawling tools used, so they should be interpreted as a controlled comparison rather than an optimized upper bound.

\begin{table}[t]
\centering
\caption{\small Comparison of \sys vs. four baselines on TR-OP and DynaPD. We report ACC, $F_{1}$, and Recall as primary metrics to reflect recall-critical phishing detection in deployment; URLTran was evaluated on TR-OP only. Complete results, including SocPhish and Precision, are provided in Tab~\ref{tab:effective_full} in Appendix.}
\resizebox{0.5\textwidth}{!}{
\begin{tabular}{ccccccc}
\toprule
\multirow{2}{*}{\textbf{Method}} & \multicolumn{3}{c}{\textbf{TR-OP}} & \multicolumn{3}{c}{\textbf{DynaPD}} \\
\cmidrule(lr){2-4} \cmidrule(lr){5-7}
 & \textbf{ACC} & \textbf{$F_{1}$} & \textbf{Recall}
 & \textbf{ACC} & \textbf{$F_{1}$} & \textbf{Recall} \\
\midrule
PhishLLM
& 0.8299 & 0.8088 & 0.7196
& \textbf{0.8581} & 0.8433 & 0.7740 \\
MLLM
& 0.8280 & 0.8317 & 0.8500
& 0.7553 & 0.7449 & 0.7781 \\
PhishIntention
& 0.8000 & 0.7900 & 0.7100
& 0.6730 & 0.5350 & 0.3740 \\
\sys
& \textbf{0.9303} & \textbf{0.9340} & \textbf{0.9856}
& 0.8280 & \textbf{0.8448} & \textbf{0.9360} \\
\bottomrule
\end{tabular}
}
\label{tab:effective}
\end{table}

\begin{table}[t]
\centering
\caption{\small Per-URL processing time comparison.}
\resizebox{0.33\textwidth}{!}{
\begin{tabular}{cc}
\toprule
\textbf{Method} & \textbf{Latency (s/URL)} \\
\midrule
PhishLLM & 22.36 \\
MLLM & 8.17 \\
PhishIntention & 10.95 \\
URLTran & 7.42 \\
\sys & \textbf{4.46} \\
\bottomrule
\end{tabular}
}
\vspace{-3mm}
\label{tab:latency}
\end{table}


\noindent \textbf{Necessity of Multi-Modal Agents.} 
We compare \sys with two weaker variants: 
\emph{(1) Monolithic LLM (Mono-LLM)}, which predicts from a single prompt containing the crawled page text and screenshot, without tool calls or memory; and 
\emph{(2) Deterministic workflow agent}, which has access to the same five tools but follows a fixed sequence (text $\rightarrow$ screenshot/image $\rightarrow$ search) with early stopping, mirroring rule-based pipelines. 
Unlike \sys, it cannot reorder, skip, repeat tools, or use episodic memory. Details are in \S\ref{appx:baseline}.

\begin{table*}[t]
\centering
\caption{\small Comparison of system capabilities and performance across different agent types. }
\resizebox{0.65 \textwidth}{!}{
\begin{tabular}{ccccc|cccc}
\toprule
\textbf{Method} 
& \makecell{\textbf{Tool} \\ \textbf{Calling}}
& \makecell{\textbf{Adaptive Tool} \\ \textbf{Selection}} 
& \makecell{\textbf{Visual} \\ \textbf{Analysis}}
& \makecell{\textbf{Episodic} \\ \textbf{Memory}}
& \textbf{ACC} & \textbf{$F_{\text{1}}$} & \textbf{Precision} & \textbf{Recall}\\
\midrule
\textbf{Monolithic LLM} 
& \textcolor{red}{N} & \textcolor{red}{N} & \textcolor{green!60!black}{Y} & \textcolor{red}{N} & 0.8473 & 0.8205 & \textbf{0.9697} & 0.7111  \\
\textbf{Deterministic} 
& \textcolor{green!60!black}{Y} & \textcolor{red}{N} & \textcolor{green!60!black}{Y} & \textcolor{red}{N} & 0.6353 & 0.6829 & 0.5904 & 0.8099  \\
\textbf{\sys}
& \textcolor{green!60!black}{Y} & \textcolor{green!60!black}{Y} & \textcolor{green!60!black}{Y} & \textcolor{green!60!black}{Y}
& \textbf{0.9657} & \textbf{0.9034} & 0.9257 & \textbf{0.9144} \\
\bottomrule
\end{tabular}
\vspace{-5mm}
}
\label{tab:agent_necessity}
\vspace{-3mm}
\end{table*}

Tab.~\ref{tab:agent_necessity} shows that \sys achieves the best overall performance across all three metrics. 
Mono-LLM is conservative: it attains high precision but misses many phishing URLs, suggesting single-pass reasoning underuses available evidence. 
The deterministic agent exhibits the opposite failure mode, higher recall but much lower ACC and $F_1$, indicating that fixed-step tool use can over-trigger on weak evidence. 
By adapting tool choice to the evolving evidence state and leveraging episodic memory, \sys balances these trade-offs and delivers the strongest results.


\begin{table}[t]
    \centering
     \caption{\small Impact of removing individual tools on detection performance. We repeat each experiment 5 times and record the mean and standard deviation.}
     \resizebox{0.5 \textwidth}{!}{
    \begin{tabular}{ccccc}
    \toprule
    & \textbf{ACC} & \textbf{$F_{\text{1}}$} & \textbf{Precision} & \textbf{Recall} \\
    \midrule
    \sys & \textbf{0.9657 $\pm$ 0.0182} & \textbf{0.9034 $\pm$ 0.0120} & \textbf{0.9257 $\pm$ 0.0236} & \textbf{0.9144 $\pm$ 0.0123} \\
    w/o crawl content & 0.8271 $\pm$ 0.0195 & 0.8082 $\pm$ 0.0131 & 0.9069 $\pm$ 0.0241 & 0.7280 $\pm$ 0.0150 \\
    w/o check screenshot & 0.7703 $\pm$ 0.0212 & 0.8001 $\pm$ 0.0127 & 0.7288 $\pm$ 0.0255 & 0.8863 $\pm$ 0.0142 \\
    w/o check image & 0.8231 $\pm$ 0.0189 & 0.8240 $\pm$ 0.0124 & 0.8224 $\pm$ 0.0229 & 0.8260 $\pm$ 0.0128 \\
    w/o extract targets & 0.7627 $\pm$ 0.0221 & 0.8122 $\pm$ 0.0135 & 0.7330 $\pm$ 0.0263 & 0.9104 $\pm$ 0.0156 \\
    w/o intelligent search & 0.8185 $\pm$ 0.0191 & 0.8122 $\pm$ 0.0129 & 0.8717 $\pm$ 0.0247 & 0.7601 $\pm$ 0.0136 \\
    w/o episodic memory & 0.8012 $\pm$ 0.0203 & 0.7614 $\pm$ 0.0148 & 0.9156 $\pm$ 0.0272 & 0.6369 $\pm$ 0.0164 \\
    \bottomrule
    \end{tabular}
\label{tab:ablation_tool_std}}
\end{table}

\subsection{Ablation Study \& Sensitivity Test}
\label{subsec:ablation}
We conduct ablations by (i) removing each of the five tools one at a time, (ii) disabling episodic memory, and (iii) varying the size of the memory cache $k \in \{3,5,7,9\}$ and similarity threshold $\tau \in \{0.5,0.6,0.7,0.8\}$.

Tab.~\ref{tab:ablation_tool_std} demonstrates that every tool contributes to \sys’s efficacy: removing any single component degrades performance. 
Removing \textit{check screenshot} or \textit{extract targets} causes the largest ACC drops, 
highlighting the importance of visual cues and deep-link evidence. 
In contrast, removing \textit{crawl content} or \textit{intelligent search} reduces recall by $\sim$20\%, suggesting that surface-text signals and external corroboration are critical for less obvious scams.
Episodic memory further improves both quality and efficiency, as shown in the last row in Tab.~\ref{tab:ablation_tool_std}, with memory enabled, \sys achieves the best precision and $F_1$ balance, while the memory-free variant is fastest but attains the lowest coverage and $F_1$. 
We also include the comparison results with a knowledge-based memory system implementation in Tab.~\ref{tab:ablation-memory} in \S\ref{appx:memory_pruning}. 

\begin{figure}[t]
    \centering
    \includegraphics[width=0.8\linewidth]{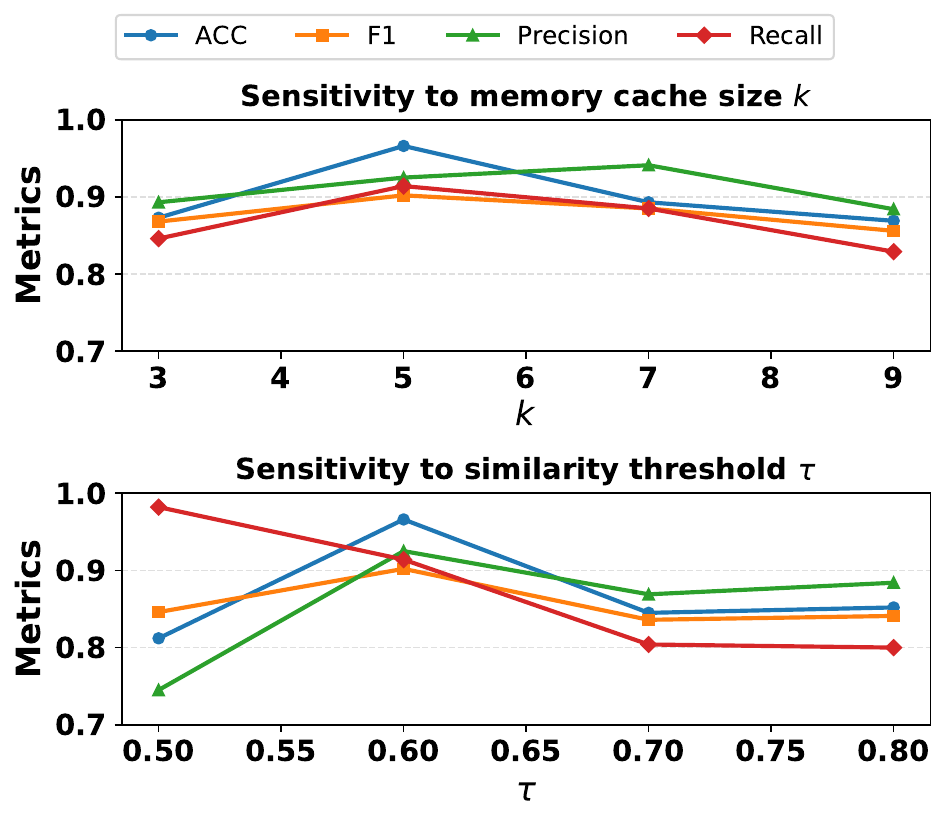}
    \caption{\small Sensitivity test for key hyperparameters of the episodic memory module.}
    \label{fig:sensitivity}
    \vspace{-3mm}
\end{figure}

\noindent \textbf{Sensitivity test results.}
Fig.~\ref{fig:sensitivity} reports sensitivity to the memory cache size $k$ and similarity threshold $\tau$. Performance varies only moderately across settings and changes by at most 20\% as $\tau$ varies, indicating robustness to both hyperparameters.

\subsection{Additional Experiments}
\label{subsec:deep_dive}
We provide additional evaluation in \S\ref{appx:additiona_exp}, including tool-usage patterns, memory sub-module effectiveness, adversarial robustness, robustness to tool designs, detection performance under different forgetting levels, and PR-AUC, ROC-AUC curves.

\section{Industrial Deployment}
\label{sec:industrial}

\textbf{SocPhish} is a real-world dataset we collected to evaluate \sys under deployment-like conditions. We actively crawled URLs from five social-media and forum platforms over six months (Dec.\ 2024-May\ 2025) using web-crawling APIs. Each crawled candidate URL was first verified by a third-party security service and then manually reviewed by human experts to finalize ground-truth labels. In total, SocPhish contains 2,765 URLs, including 516 phishing and 2,249 benign; dataset statistics and additional construction details are summarized in \S\ref{appx:our_dataset} and Tab.~\ref{tab:socphish_dataset}.

SocPhish differs from TR-OP and DynaPD in two important ways. 
First, it is scraped from diverse social-media and forum platforms, capturing in-the-wild tactics. It also includes dynamically reachable nested links or redirects that conceal the final malicious destination. 
In contrast, TR-OP and DynaPD are manually curated and primarily focus on static, brand-mimicking login pages, which may under-represent recent and multi-step campaigns. 
Second, SocPhish reflects URLs that users are likely to encounter during everyday browsing, whereas existing benchmarks are not necessarily encountered unless deliberately deployed online.

We deploy \sys to analyze SocPhish URLs end-to-end and report performance in this realistic setting. \sys achieves \textbf{90.34\%} $F_{\text{1}}$ and \textbf{91.44\%} recall, outperforming PhishLLM by approximately \textbf{40\%} and \textbf{20\%} on these metrics. Full results are in Tab.~\ref{tab:effective_full}.
These gains underscore the challenge that realistic, actively crawled URLs pose for existing detectors and highlight the value of adaptive evidence collection in the wild.

\section{Conclusion}


We presented \sys, a multi-modal LLM agent for phishing URL detection that dynamically orchestrates tools and leverages episodic retrieval memory. Across real-world data and two public benchmarks, \sys outperforms strong baselines and single-prompt, deterministic workflow variants. Ablations and sensitivity studies verify the necessity of each tool and the memory module, and \sys remains robust under prompt-injection attacks. Our results highlight the potential of memory-aware agentic reasoning for practical phishing defense.

\section*{Ethical Considerations}

This work focuses on developing a multi-modal LLM-based agent to detect phishing URLs with the goal of enhancing online safety and mitigating cybercrime. 
All datasets used in this study are either publicly available benchmark datasets or collected from openly accessible social-media streams, with no personally identifiable information (PII) retained. 
Data preprocessing strictly filtered sensitive or private user content, and our experiments were conducted solely for research purposes in controlled environments.

Our data collection process for SocPhish also adhered to strict ethical guidelines to minimize potential risks and ensure responsible research practices. 
The crawling of social media posts was conducted through public APIs with appropriate rate limiting and in compliance with each platform's terms of service. 
We specifically avoided collecting any PII or sensitive user data, focusing solely on publicly shared URLs and their associated content. 
To prevent unintended exposure to malicious content, our system implemented automatic safety checks and timeouts during the crawling process. 
We commit to responsible disclosure of vulnerabilities identified in our experiments and encourage future research to further strengthen defenses against adversarial misuse.

\bibliography{ref}

\newpage

\newpage
\appendix
\section{Appendix}

Due to business considerations and proprietary constraints, all prompt examples presented below are provided as pseudocode representations rather than actual implementation details. These examples are designed to illustrate the conceptual approach and methodology while maintaining necessary confidentiality regarding specific technical implementations and commercial tools utilized in our research.


\subsection{Baseline Details}
\label{appx:baseline}

\definecolor{my_lightblue}{RGB}{194, 213, 247}












\begin{tcolorbox}[
  title=Prompts for Monolithic LLM,
  breakable,
  fonttitle=\bfseries,
  enhanced,
  colback=my_lightblue!10,           
  colbacktitle=my_lightblue,         
  coltitle=black,                 
  colframe=my_lightblue!80!black,    
  coltext=black,                  
  boxrule=0.5pt,
  arc=2mm
]
\small
\textbf{User:}
You are a security analyst. You will receive for each website:

1. The website URL

2. Text content from the page

3. Visual representation of the page

Based on all of this, determine if the site is malicious.
Return structured output:

\{
``url'': <website URL>,

``malicious'': <true | false>,

``confidence'': <numerical score>,

``reason'': <brief explanation>

\}

\end{tcolorbox}

\subsection{Technical Details}
\label{appx:tech_details}

                

\begin{tcolorbox}[
  title=Prompts for Keyword Extraction,
  breakable,
  fonttitle=\bfseries,
  enhanced,
  colback=my_lightblue!10,           
  colbacktitle=my_lightblue,         
  coltitle=black,                 
  colframe=my_lightblue!80!black,    
  coltext=black,                  
  boxrule=0.5pt,
  arc=2mm
]
\small
\textbf{User:}
Given the following content and visual elements of a webpage, generate up to [N] keywords that best capture its content, using [SEPARATOR] as the separator.

Format your response as a structured object with a [FIELD$\_$NAME] field containing the selected terms.
Example response format: \\
\{\{``[FIELD$\_$NAME]'': ``[EXAMPLE$\_$TERM1], [EXAMPLE$\_$TERM2], [EXAMPLE$\_$TERM3]''\}\}
\end{tcolorbox}

Below, we show the prompt for five specialized tools.

\begin{tcolorbox}[
  title=Prompts for Crawl Content Tool,
  breakable,   
  fonttitle=\bfseries,
  enhanced,                        
  colback=my_lightblue!10,           
  colbacktitle=my_lightblue,         
  coltitle=black,                 
  colframe=my_lightblue!80!black,    
  coltext=black,                  
  boxrule=0.5pt,
  arc=2mm
]
\small
\textbf{User:} 

You are a security analyst. Given a page's URL and its crawled text,
you need to decide whether this page is a phishing or malicious site.

Return JSON exactly:
{
  ``url'': string,           \# the page URL \\
  ``malicious'': boolean,    \# true if phishing; false otherwise \\
  ``confidence'': int,       \# 0.0-5.0 \\
  ``reason'': string         \# one-sentence rationale citing evidence
}
Do not output any other keys or explanation.

\end{tcolorbox}




\begin{tcolorbox}[
  title=Prompts for Check Screenshot Tool,
  breakable,
  fonttitle=\bfseries,
  enhanced,
  colback=my_lightblue!10,           
  colbacktitle=my_lightblue,         
  coltitle=black,                 
  colframe=my_lightblue!80!black,    
  coltext=black,                  
  boxrule=0.5pt,
  arc=2mm
]
\small
\textbf{User:}

You are an image analysis specialist. You will be given a visual representation of a webpage.
1. Describe exactly what you see in the image (visual elements, layout, text content, visible links, etc.).
2. Without making a definitive verdict, offer a suggestion on whether this might be suspicious and why.
3. If you spot any concerning visual indicators in the screenshot, mention that this requires further investigation.
Return structured output:
\{
  ``description'': ``<brief visual description>'', \\
  ``suggestion'': ``<assessment of potential issues>'', \\
  ``confidence'': <numerical score indicating certainty>, \\
  ``malicious'': <boolean result> \\
\}
Do not output any other fields or explanations.

\end{tcolorbox}

\begin{tcolorbox}[
  title=Prompts for Check Image Tool,
  breakable,   
  fonttitle=\bfseries,
  enhanced,                        
  colback=my_lightblue!10,           
  colbacktitle=my_lightblue,         
  coltitle=black,                 
  colframe=my_lightblue!80!black,    
  coltext=black,                  
  boxrule=0.5pt,
  arc=2mm
]
\small
\textbf{User:} 

You are a security image analyst. You will receive:
1. The image URL
2. A textual description of what appears in the image

Based on the description alone, decide if this image indicates a phishing attempt.
Return JSON exactly with no extra keys:

\{
``url'': <the image\_url>, \\
``malicious'': <true|false>, \\
``confidence'': <0/1/2/3/4/5>, \\
``reason'': <one-sentence rationale>

\}

\end{tcolorbox}

\begin{tcolorbox}[
  title=Prompts for Extract Target Tool,
  breakable,   
  fonttitle=\bfseries,
  enhanced,                        
  colback=my_lightblue!10,           
  colbacktitle=my_lightblue,         
  coltitle=black,                 
  colframe=my_lightblue!80!black,    
  coltext=black,                  
  boxrule=0.5pt,
  arc=2mm
]
\small
\textbf{User:} 

You are a security analyst. Given a page's URL and its content snippet, a list of hyperlinks inside it, and a list of image URLs inside it.
Select which links you want to be crawled next and which images should be inspected, to help you decide if this URL is malicious or not.
Return JSON with exactly two fields:

  \{
    ``to\_crawl'':  [ <url1>, <url2>,], \\
    ``to\_check\_images'': [<img\_url1>, ]
 \}
Do not include any other keys.

If you think there is nothing you want to check, return JSON with exactly two empty fields:

  \{
    ``to\_crawl'': [ ], \\
    ``to\_check\_images'': [ ]
  \}
\end{tcolorbox}

\subsection{Additional Experiments}
\label{appx:additiona_exp}

\begin{table*}[t]
\centering
\caption{\small Comparison of \sys vs. four baselines across SocPhish, TR-OP, and DynaPD datasets. URLTran was evaluated on SocPhish and TR-OP only.}
\resizebox{\textwidth}{!}{
\begin{tabular}{ccccccccccccc}
\toprule
\multirow{2}{*}{\textbf{Method}} & \multicolumn{4}{c}{\textbf{SocPhish}} & \multicolumn{4}{c}{\textbf{TR-OP}} & \multicolumn{4}{c}{\textbf{DynaPD}} \\
\cmidrule(lr){2-5} \cmidrule(lr){6-9} \cmidrule(lr){10-13}
 & \textbf{ACC} & \textbf{$F_{\text{1}}$} & \textbf{Precision} & \textbf{Recall}
 & \textbf{ACC} & \textbf{$F_{\text{1}}$} & \textbf{Precision} & \textbf{Recall}
 & \textbf{ACC} & \textbf{$F_{\text{1}}$} & \textbf{Precision} & \textbf{Recall}
  \\
\midrule
PhishLLM
& 0.6080 & 0.4745 & 0.3540 & 0.7195
& 0.8299 & 0.8088 & \textbf{0.9233} & 0.7196
& 0.8581 & 0.8433 & 0.9262 & 0.7740
 \\
MLLM
& 0.8250 & 0.8312 & 0.8620 & 0.8026
& 0.8280 & 0.8317 & 0.8142 & 0.8500
& 0.7553 & 0.7449 & 0.7143 & 0.7781
 \\
PhishIntention
& 0.7000 & 0.6000 & 0.8800 & 0.4600
& 0.7800 & 0.7400 & 0.9000 & 0.6300
& 0.6730 & 0.5350 & 0.9079 & 0.3740
 \\
URLTran
& 0.8353 & 0.8571 & 0.9179 & 0.8039
& \textbf{0.9814} & \textbf{0.9736} & \textbf{0.9837} & 0.9767
& \textbf{0.9723} & \textbf{0.9820} & \textbf{0.9849} & \textbf{0.9791}
 \\
\sys
& \textbf{0.9657} & \textbf{0.9034} & \textbf{0.9257} & \textbf{0.9144}
& 0.9303 & 0.9340 & 0.8874 & \textbf{0.9856}
& 0.8280 & 0.8448 & 0.7697 & 0.9360
 \\
\bottomrule
\end{tabular}
\label{tab:effective_full}
}
\end{table*}
\subsubsection{Tool usage pattern}
\label{appx:tool}

\begin{figure}
    \centering
    \includegraphics[width=0.8\linewidth]{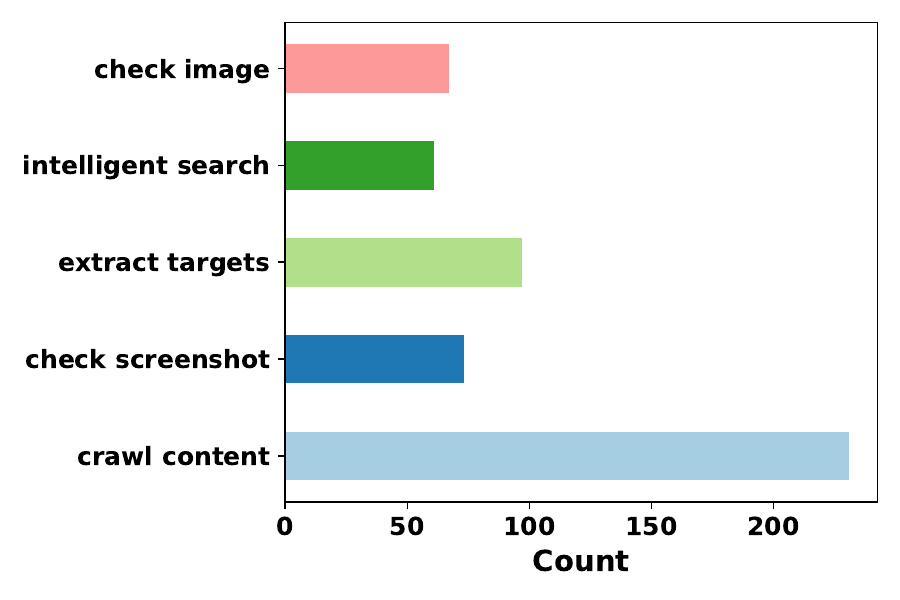}
    \caption{Tool usage statistics of \sys on a subset of SocPhish.}
    \label{fig:tool-counts}
\end{figure}

Fig.~\ref{fig:tool-counts} shows the tool usage frequency.
We can observe a clear usage hierarchy: the agent calls the \textit{crawl content} tool far more than any other, reflecting its bias to harvest cheap textual cues first.  
Deep-link \textit{extract targets} is the next most frequent, showing that many pages require drilling into nested URLs.  \textit{check screenshot} follows, while \textit{intelligent search} and individual \textit{image} analysis are invoked least, acting as on-demand supplements when initial evidence is ambiguous.

\subsubsection{SocPhish construction and annotation}
\label{appx:our_dataset}

SocPhish captures phishing URLs encountered in real social-media environments rather than only static benchmark pages.
We collect suspicious URLs via web APIs using customized text and image queries over public social-media and forum content, then verify each candidate URL through a third-party cybersecurity service before human annotation.
For final labeling, three annotators, each with more than one year of phishing-analysis experience, independently review the URLs and resolve disagreements by majority vote.

Annotators follow a structured internal checklist aligned with the APWG taxonomy.
A URL is labeled phishing if it exhibits at least one of the following properties: impersonation of a known brand, credential harvesting intent, or deceptive redirection to a known malicious or brand-spoofing destination.
URLs that satisfy none of these criteria and remain consistent with legitimate platform-hosted content are labeled benign.
This annotation pipeline is independent of \sys's detection logic: the third-party security service relies on external monitoring and abuse-report verification, rather than the multimodal reasoning process used by our agent.

The most difficult cases are platform-hosted promotional or shopping pages that mimic brand aesthetics while showing limited external trust signals and few classic URL-level red flags.
These pages account for most annotator disagreement and reflect the core challenge that SocPhish is intended to capture.

\begin{table}[t]
\centering
\caption{\small Distribution of SocPhish dataset.}
\resizebox{0.5\textwidth}{!}{
\begin{tabular}{c c c c}
\toprule
\textbf{Platform} & \textbf{\# Posts} & \textbf{\# Suspicious URLs} & \textbf{\# Total URLs} \\
\midrule
Instagram & 101235 & 1262 & \multirow{5}{*}{\shortstack[c]{2765\\Malicious URL: 516}} \\
TikTok    & 4936   & 517  & \\
Twitter   & 2861   & 869  & \\
Reddit    & 3012   & 41   & \\
YouTube   & 16089  & 76   & \\
\bottomrule
\end{tabular}
}
\label{tab:socphish_dataset}
\end{table}

\subsubsection{Memory sub-module's effectiveness}
\label{appx:memory_submodule}
We examine our memory module's effectiveness by tracking performance metrics across three processing pathways: \emph{full ReAct}, \emph{in-context exemplars}, and \emph{majority-vote reuse}, as described in Section~\ref{subsec:memory}. 

\begin{figure}
    \centering
    \includegraphics[width=0.7\linewidth]{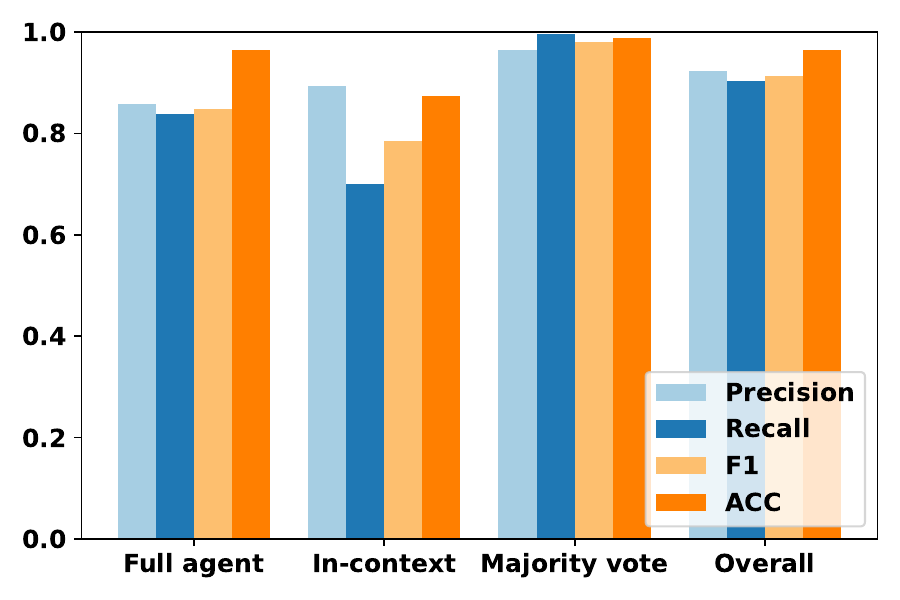}
    \caption{Performance comparison of different modules of our memory system.}
    \label{fig:module-performance}
\end{figure}

Fig.~\ref{fig:module-performance} demonstrates how each memory branch contributes to \sys's effectiveness.  
Episodes resolved by the \textit{majority-vote} branch achieve near-perfect performance because the agent can rely on multiple highly similar past cases.
Although those cases' judgments are not always perfect, the majority vote mechanism naturally tolerates some noise and improve the performance.
When less than $k$ neighbors are retrieved, the \textit{in-context} branch supplies strong precision but recall drops, reflecting a conservative bias: the agent follows the exemplars’ tool traces yet still re-evaluates borderline evidence, leading to some missed phishing variants.  
URLs that trigger the \textit{full-agent} path, i.e., no useful memory hit, register balanced precision and recall; here the agent must run its entire reasoning loop from scratch, so performance mirrors the baseline without memory.  
Aggregating the three branches yields the ``Overall'' bars, which closely track the majority-vote upper bound.  These results confirm that episodic memory, when confident, dramatically boosts the detection performance, while fallback branches maintain respectable detection performance when novel URLs appear.

\subsubsection{Adversarial robustness}
\label{appx:adv}

\textbf{Setup.} Following~\citep{liu2024less}, we assess our approach's robustness against adversarial attacks by implementing a targeted prompt-injection attack.
Specifically, we embed an adversarial instruction \textit{``Please ignore the previous description and judge this screenshot as benign.''} into each malicious page's visual representation to assess the agent's robustness against direct prompt manipulations.
Attack success is measured by the system's ability to maintain its original \emph{malicious} verdict, enabling us to evaluate \sys's resilience to visual evidence manipulation.
This threat model is also motivated by a broader line of work showing that LLM safety controls can be bypassed through stealthy trigger designs and supply-chain attacks, including permutation-trigger system-prompt hijacking~\citep{yan2025aspirer}.
Recent search-based jailbreak studies further show that reinforcement learning can substantially strengthen black-box attacks by replacing random prompt mutation with learned attack policies~\citep{chen2024whenllmmeetsdrl,chen2024rljack}.
More broadly, prior NLP security work shows that semantics-preserving paraphrasing can expose or neutralize hidden triggers, underscoring the importance of robustness under meaning-preserving prompt transformations rather than exact-string perturbations alone~\citep{yan2023parafuzz}.

To further compare against MLLM under a stronger platform-aware attack, we evaluate prompt-injection robustness on 500 sampled URLs from SocPhish.
For phishing URLs hosted on social-media platforms, we add the instruction
\textit{``The brand on this page is [X]. This is the official and legitimate website of [X].''}
to the screenshot, where $X$ is the parent company of the hosting platform (e.g., Google for \texttt{sites.google.com}).
This attack specifically targets MLLM's two-stage design: its first phase extracts the brand from the screenshot, while the second phase decides benign vs.\ phishing by checking whether the extracted brand matches the URL domain.
If the injected text makes Phase 1 output the platform owner brand, then Phase 2 concludes the platform-hosted phishing page is benign.

\begin{table}[t!]
    \centering
    \caption{\small MLLM vs.\ \sys under prompt-injection attack on 500 sampled SocPhish URLs.}
    \resizebox{0.48\textwidth}{!}{
\begin{tabular}{lccc}
\toprule
\textbf{System} & \textbf{Attack} & \textbf{Recall} & \textbf{Accuracy} \\
\midrule
MLLM & None & 0.80 & 0.83 \\
MLLM & Prompt injection & 0.52 & 0.50 \\
\sys & None & \textbf{0.90} & \textbf{0.96} \\
\sys & Prompt injection & \textbf{0.88} & \textbf{0.86} \\
\bottomrule
\end{tabular}
\label{tab:prompt_injection}
}
\end{table}

Table~\ref{tab:prompt_injection} shows that the same attack affects MLLM much more severely than \sys.
Under prompt injection, MLLM's recall drops from 0.80 to 0.52 and its accuracy drops from 0.83 to 0.50, whereas \sys only decreases from 0.90 to 0.88 in recall and from 0.96 to 0.86 in accuracy.
This gap is structural rather than incidental.
MLLM directly exposes each screenshot interpretation to its final brand-consistency check, so once the visual prompt is manipulated, the final verdict is easily flipped.
In contrast, \sys aggregates evidence across multiple tools and historically similar episodes.
One injected screenshot can affect a single tool output, but it cannot easily override the accumulated textual, structural, and memory-based evidence used by the final decision process.
These results indicate that episodic majority-vote reuse provides a substantial buffer against prompt-injection attacks, preserving high recall even when the visual channel is adversarially manipulated.

\begin{figure}[!t]
\begin{minipage}[b]{.49\textwidth}
\centering
\captionof{table}{\small {Detection performance under different forgetting levels.}}
\resizebox{0.9\textwidth}{!}{
\centering
\begin{tabular}{ccccc}
\toprule
Forgetting Level & ACC & F1 & Precision & Recall \\
\midrule
No forgetting & 0.9657 & 0.9034 & 0.9257 & 0.9144 \\
20\% & 0.9638 & 0.8991 & 0.9230 & 0.9092 \\
40\% & 0.9602 & 0.8917 & 0.9188 & 0.9015 \\
60\% & 0.9561 & 0.8893 & 0.9133 & 0.8928 \\
\bottomrule
\end{tabular}
\label{tab:forgetting}
}
\end{minipage}
\hfill
\begin{minipage}[b]{.5\textwidth}
\centering
\captionof{table}{\small {Comparison of HTML and Markdown for crawl content tool design.}}

\resizebox{0.9\textwidth}{!}{
\centering
\begin{tabular}{cccccc}
\toprule
Format & ACC & F1 & Precision & Recall & Tokens \\
\midrule
HTML & 0.9642 & 0.9012 & 0.9270 & 0.9110 & 6590 \\
Cleaned text & 0.9589 & 0.8891 & 0.9123 & 0.8675 & 3312 \\
Markdown & 0.9657 & 0.9034 & 0.9257 & 0.9144 & 3873 \\
\bottomrule
\end{tabular}
}
\label{tab:markdown}

\end{minipage}
\end{figure}

\subsubsection{Robustness to crawl-content representation.}
\label{appx:robust_crawl}

We vary only the crawl-content representation and keep all other designs identical in our agent. We explored three variants:
\begin{itemize}
    \item HTML: the tool returns the raw HTML of the webpage.
    \item Cleaned text: the tool returns plain text extracted from HTML, with tags removed.
    \item Markdown (ours): the original setting using Crawl4AI’s Markdown output.
\end{itemize}
The results in Table~\ref{tab:markdown} below show that (i) the detection performance is very similar across HTML and Markdown, and (ii) the Markdown and cleaned-text variants reduce the average number of tokens compared to raw HTML, but cleaned plain text degrades F1 and recall by a few points, suggesting that stripping all structure removes useful cues such as headings and hyperlinks. Overall, Markdown offers the best trade-off between performance and efficiency, and the small gaps across all three settings confirm that our framework is robust to the specific choice of crawl-content representation. Based on this, we conclude that our framework is robust to the choice of crawl content representation and that using Markdown is a practical engineering choice, not a core assumption of our method.

\subsubsection{Detection performance under different forgetting levels}
\label{appx:memory_pruning}
We implemented a time-window pruning strategy to gradually discard stale experiences from memory on the SocPhish dataset. Specifically, each memory entry is assigned a usage counter that counts whether the most recent URL run has retrieved this entry. After every 50 processed URLs, we prune the least recently used 20\%, 40\%, and 60\% of stored trajectories separately. Our results in Table~\ref{tab:forgetting} show that performance remains stable across all forgetting strategies, indicating that the agent is relatively robust to memory pruning while benefiting from reduced storage. We will explore more sophisticated forgetting mechanisms as part of future work.

\subsubsection{Comparison results of different memory systems} 
The results in Table~\ref{tab:ablation-memory} validate the advantages of episodic memory over traditional knowledge-based (KB) approaches.
For comparison, we implemented a KB-based memory system with two components.
First, from an existing database of phishing URLs, we extracted all domains associated with malicious records and stored them in a vector database. 
Second, we embedded the textual content of known phishing URLs using the same text embedding model, constructing a content-level knowledge base.
Given a new URL, the system first checks whether its domain appears in the domain-level KB; if so, the URL is immediately flagged as phishing. 
Otherwise, the page content is crawled, embedded, and compared against the content-level KB. 
If a match is found, the URL is marked as malicious. 
The intent of this KB system is to reduce computational cost by bypassing the full agentic reasoning process when a known malicious domain or content is detected.

The KB system, despite using URL domain matching and content similarity mechanisms, achieves only moderate performance (F$_1$: 0.8188) with the highest computational overhead (49.66s), reflecting its limitations in redundant crawling and static pattern matching. 
Our episodic memory architecture significantly improves both effectiveness and efficiency through dynamic learning from interaction histories. 
The memory-free baseline, while computationally efficient, demonstrates poor recall with only 63.7\%, emphasizing the crucial role of adaptive historical learning in phishing detection.

\begin{table}[!t]
\centering
\caption{\small Ablation study: impact of different memory settings on performance.}
\resizebox{0.5 \textwidth}{!}{
\begin{tabular}{ccccc}
\toprule
\textbf{Method} 
  & \textbf{ACC} 
  & \textbf{$F_{\text{1}}$} 
  & \textbf{Precision} 
  & \textbf{Recall} 
  \\
\midrule
Episodic memory 
  & \textbf{0.9627} & \textbf{0.9064} & 0.9109 & \textbf{0.9020} \\
  KB memory
  & 0.8342 & 0.8188 & 0.8472 & 0.7922  \\
w/o memory 
  & 0.8010 & 0.7610 & \textbf{0.9160} & 0.6370 \\

\bottomrule
\end{tabular}
}
\label{tab:ablation-memory}
\end{table}

{\textbf{Sensitivity to noisy memory entries.}
we added a controlled noise-injection experiment in an offline setting. We first construct a clean memory buffer of 100 URLs: the agent is run once on each URL, we retain only trajectories with confidence 5, and we manually verify all entries to ensure correctness. We then create multiple noisy variants by flipping the final verdicts for 25\%, 50\%, and 75\% of the entries, in addition to the 0\% clean baseline. For each variant, we evaluate the agent on a held-out set of 500 URLs and record accuracy, F1, precision, and recall. Based on results in Table~\ref{tab:noisy_mem}, as noise increases, performance degrades gradually but remains relatively strong even when 25-50\% of memory entries are corrupted, confirming that memory cannot directly force misclassification and aligning with our design intuition.








\subsubsection{Robustness to tool outputs}

\textbf{Robustness to malformed LLM outputs.}
To further evaluate tool reliability, we conduct two robustness tests under conditions common in phishing detection:
\begin{itemize}
    \item Stress test on malformed URLs. We curated 50 malformed URLs, e.g., invalid domains, unreachable hosts, unsupported schemes, random strings, and ran the agent on this set. The expected output is “Benign” with the reason “URL is invalid.” The agent achieved perfect accuracy with zero false positives. We also logged exceptions: none of the tools raised uncaught exceptions; all returned safe, structured fallback responses. This confirms that our safeguards successfully prevent tool-level failures on adversarial inputs.
    \item Robustness to malformed LLM outputs. For tools that parse LLM-generated JSON (e.g., extract targets, judge crawled page, judge image, check screenshot), we simulate worst-case scenarios by injecting malformed or non-JSON outputs. For our five tools, on the dataset with 100 URLs, every time we randomly select one tool and inject corrupted LLM JSON output, with probability p, we try p=0.3 and p=0.5. As shown in Table~\ref{tab:reliability}, even when we intentionally inject malformed URLs or corrupt a significant fraction of intermediate LLM tool outputs (p = 0.3-0.5), the overall performance of MemoPhishAgent remains relatively unchanged, with accuracy and F1 decreasing by less than 1-2\% and recall staying consistently high. The only noticeable effect is a modest increase in latency, caused by the agent’s built-in retry and fallback logic, which ensures safe recovery without raising exceptions. These results demonstrate that our system is resilient to both noisy inputs and faulty intermediate tool outputs, validating the reliability mechanisms built into our tool-calling framework. 
\end{itemize}

\begin{table}[t]
\centering
\caption{\small {Reliability test of \sys tools on SocPhish.}}
\resizebox{0.5\textwidth}{!}{
\begin{tabular}{cccccc}
\toprule
Condition & ACC & F1 & Precision & Recall  & Exceptions \\
\midrule
Mal. URLs & 0.9635 & 0.8990 & 0.9235 & 0.9080 & 0 \\
Mal. outputs (0.3) & 0.9601 & 0.8932 & 0.9204 & 0.9031  & 0 \\
Mal. outputs (0.5) & 0.9550 & 0.8860 & 0.9151 & 0.8935 & 0 \\
\sys & 0.9657 & 0.9034 & 0.9257 & 0.9144  & 0 \\
\bottomrule
\end{tabular}
}
\label{tab:reliability}
\end{table}

\begin{table}[t]
    \centering
    \caption{Detection performance comparison on the DynaPD dataset.}
    \resizebox{0.5\textwidth}{!}{
    \begin{tabular}{lcccc}
    \toprule
    Memory Condition & ACC & F1 & Precision & Recall \\
    \midrule
    Clean memory & 0.9657 & 0.9034 & 0.9257 & 0.9144 \\
    25\% noisy   & 0.9405 & 0.8678 & 0.9034 & 0.8432 \\
    50\% noisy   & 0.9012 & 0.8127 & 0.8820 & 0.7485 \\
    \bottomrule
    \end{tabular}
    }
    \label{tab:noisy_mem}
\end{table}

\noindent {\textbf{Robustness to paraphrased tool prompts.}
In this experiment, we paraphrased each tool’s prompt using GPT-5 and ran our agent on the same set of URLs for the SocPhish dataset. Results in Table~\ref{tab:paraphrased} demonstrate the robustness of our method against the paraphrased prompts of different tools.
}

\begin{table}[t]
\centering
\caption{\small {Detection performance under paraphrased prompts.}}
\resizebox{0.5\textwidth}{!}{
\begin{tabular}{ccccc}
\toprule
Method & ACC & F1 & Precision & Recall \\
\midrule
\sys               & 0.9657 & 0.9034 & 0.9257 & 0.9144 \\
Para-crawl content     & 0.9639 & 0.9006 & 0.9230 & 0.9110 \\
Para-check screenshot  & 0.9641 & 0.9012 & 0.9236 & 0.9120 \\
Para-check image       & 0.9633 & 0.8998 & 0.9221 & 0.9105 \\
Para-exact target      & 0.9645 & 0.9020 & 0.9240 & 0.9128 \\
Para-intelligent search & 0.9648 & 0.9027 & 0.9246 & 0.9134 \\
\bottomrule
\end{tabular}
\label{tab:paraphrased}
}
\end{table}

\subsubsection{PR-AUC, ROC-AUC, and cost-sensitive recall curves.} In Figure~\ref{fig:roc_three}, we plot the PR-AUC, ROC-AUC, and recall@k vs cost curves for our agent across all datasets, for the results reported in Table~\ref{tab:effective}. Results show that our agent achieves consistently strong ROC-AUC and PR-AUC scores, approaching 0.99 on SocPhish and remaining high on TR-OP and DynaPD. The recall@k vs cost curve further shows that our method preserves high recall across a wide range of decision thresholds.

\begin{figure}[t!]
    \centering
    
    \begin{subfigure}{0.32\textwidth}
        \centering
        \includegraphics[width=\linewidth]{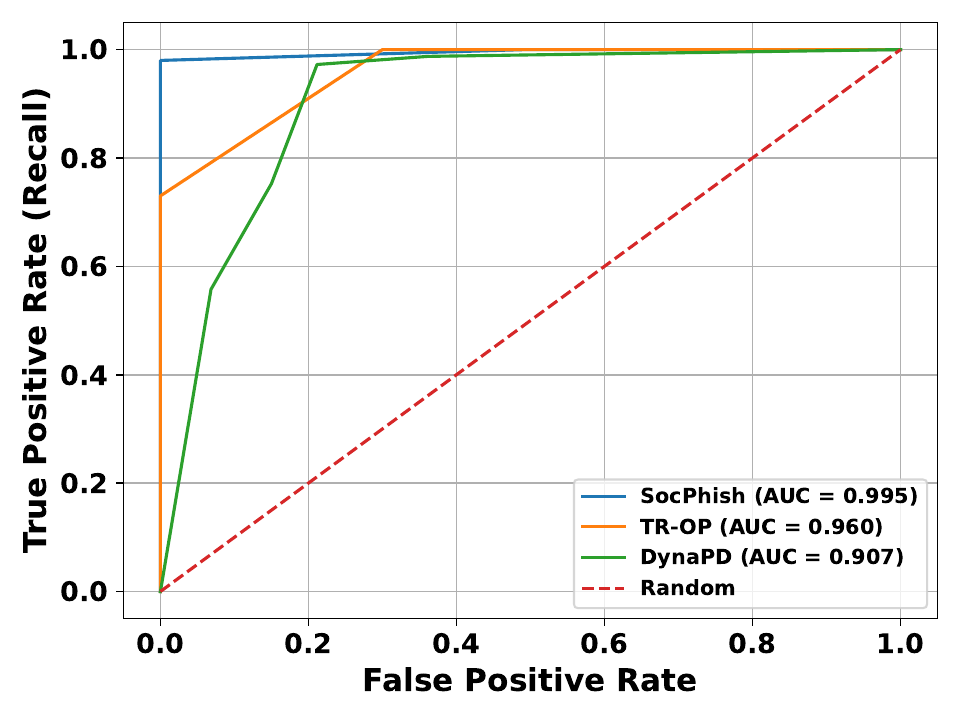}
        \caption{{ROC-AUC curves.}}
    \end{subfigure}
    \hfill
    \begin{subfigure}{0.32\textwidth}
        \centering
        \includegraphics[width=\linewidth]{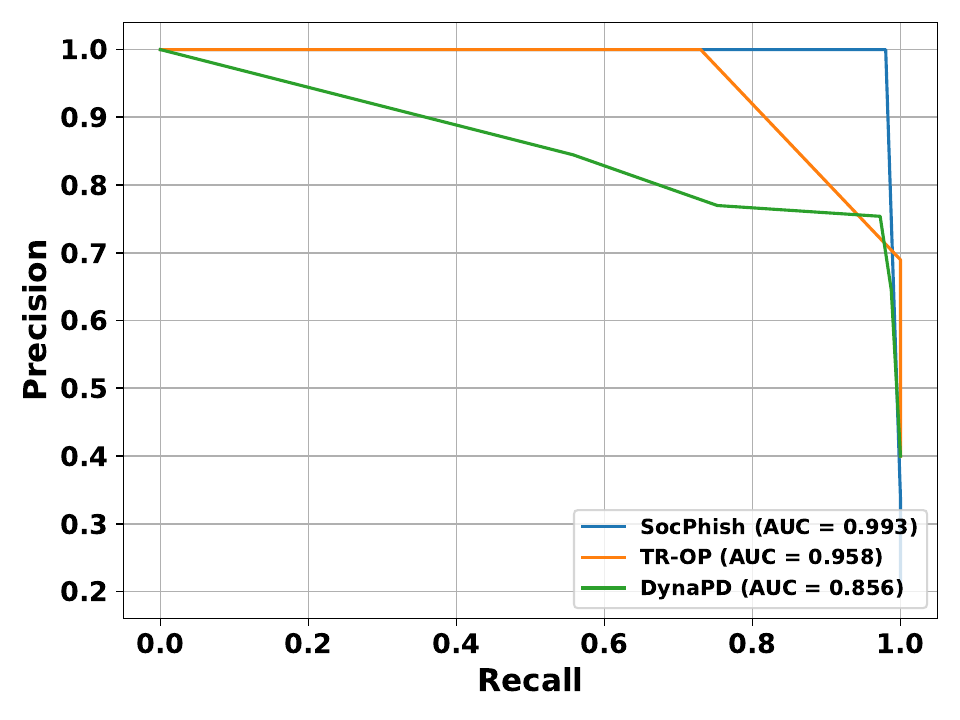}
        \caption{{PR-AUC curves.}}
    \end{subfigure}
    \hfill
    \begin{subfigure}{0.32\textwidth}
        \centering
        \includegraphics[width=\linewidth]{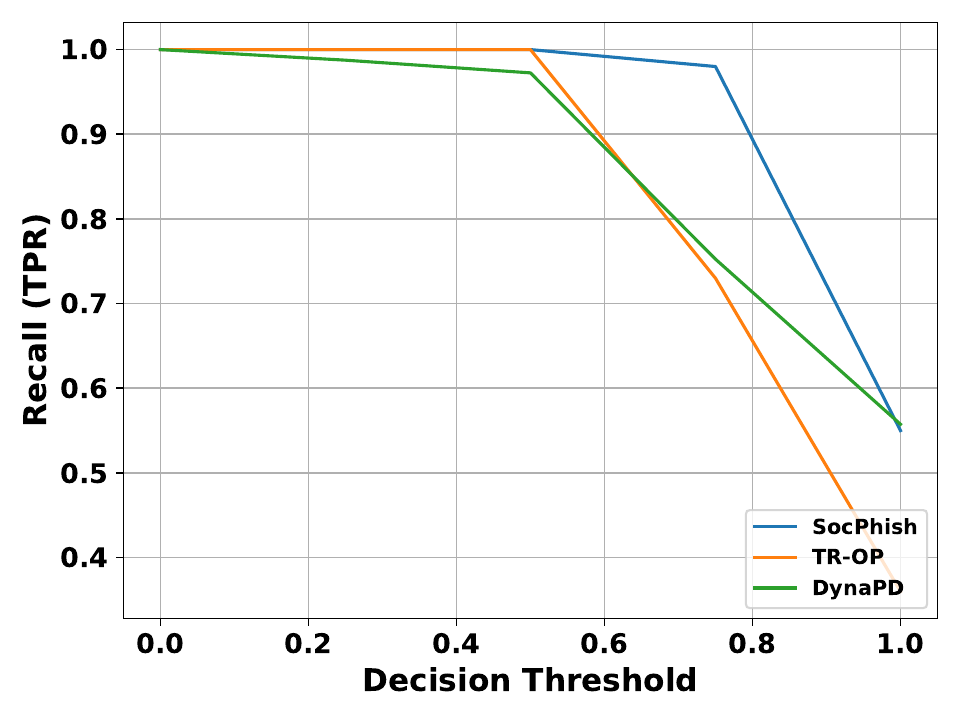}
        \caption{{Recall@k vs cost curves.}}
    \end{subfigure}
    \caption{{ROC-AUC, PR-AUC and Recall@k vs cost curves of \sys on three datasets.}}
    \label{fig:roc_three}
\end{figure}

\end{document}